\pdfoutput=1
\documentclass
[aps,prd,twocolumn,showpacs,superscriptaddress,amsfont,graphicx,nofootinbib]{revtex4}
\usepackage{color,graphicx,epsfig}
\usepackage{ifpdf}
\usepackage{amsmath}
\usepackage{bm}
\usepackage{color}
\usepackage[english]{babel}
\usepackage{graphicx}
\usepackage{amsfonts}
\usepackage{amssymb}
\usepackage{braket}

\definecolor{Red}{rgb}{1.,0.,0.}

\newcommand{\op}{{\cal O}}
\newcommand{\nn}{\nonumber}
\begin{document}

\title{New physics in $t\to b W$ decay at next-to-leading order in QCD }

\author{Jure Drobnak}
\email[Electronic address:]{jure.drobnak@ijs.si}
\affiliation{J. Stefan Institute, Jamova 39, P. O. Box 3000, 1001  Ljubljana, Slovenia} 

\author{Svjetlana Fajfer} 
\email[Electronic address:]{svjetlana.fajfer@ijs.si} 
\affiliation{J. Stefan Institute, Jamova 39, P. O. Box 3000, 1001
  Ljubljana, Slovenia}
\affiliation{Department of Physics,
  University of Ljubljana, Jadranska 19, 1000 Ljubljana, Slovenia}

\author{Jernej F. Kamenik}
\email[Electronic address:]{jernej.kamenik@ijs.si} 
\affiliation{J. Stefan Institute, Jamova 39, P. O. Box 3000, 1001  Ljubljana, Slovenia}

\date{\today}

\begin{abstract}
We consider contributions of non-standard $tbW$ effective operators to the decay of an unpolarized top quark into a bottom quark and a $W$ gauge boson at next-to-leading order in QCD.  We find that $\mathcal O_{LR}\equiv  \bar b_L \sigma_{\mu\nu} t_R W^{\mu\nu}$ contribution to the transverse-plus $W$ helicity fraction ($\mathcal F_+$) is significantly enhanced compared to the leading order result at non-vanishing bottom quark mass. Nonetheless, presently the most sensitive observable to direct $\mathcal O_{LR}$ contributions is the longitudinal $W$ helicity fraction $\mathcal F_L$.  In particular, the most recent CDF measurement of $\mathcal F_L$ already provides the most stringent upper bound on $\mathcal O_{LR}$ contributions, even when compared with indirect bounds from the rare decay $B \to X_s \gamma$.
\end{abstract}
\maketitle

\section{Introduction}
There has been a continuing interest in the measurement of helicity fractions of the $W$ boson from top quark decays by the CDF and D0 collaborations at the Tevatron. Presently, the most precise values are provided by the CDF collaboration \cite{Aaltonen:2010ha}
\begin{subequations}
\begin{align}
\mathcal F_L  &\equiv \Gamma^{L}/\Gamma =  0.88 \pm 0.11(\text{stat.}) \pm 0.06(\text{sys.}) \label{e1} \,,\\
\mathcal F_+  &\equiv \Gamma^{+}/\Gamma = -0.15 \pm 0.07(\text{stat.}) \pm 0.06(\text{sys.}) \,, 
\end{align}
\end{subequations}
where $\Gamma^L$ and $\Gamma^+$ denote the rates into the longitudinal and transverse-plus polarization state of the $W$ boson, while $\Gamma$ is the total rate. Note that the central CDF value of $\mathcal F_+$  lies outside of the physical region. In the near future, the large $t\bar t$ production cross section at the LHC is expected to provide an opportunity to study $tbW$ interactions at the percent level accuracy~\cite{AguilarSaavedra:2007rs}. It is therefore important to carefully evaluate and understand the implications of such measurements within the Standard Model (SM) and beyond.

In the SM, simple helicity considerations show that $\mathcal F_+$ vanishes at the Born term level in the $m_b = 0$ limit. A non-vanishing transverse-plus rate could arise from i) $m_b\neq 0$  effects, ii) $\mathcal O(\alpha_s)$ radiative corrections due to gluon emission\footnote{Electroweak corrections also contribute, but turn out to be much smaller~\cite{Do:2002ky}.}, or from iii) non-SM $tbW$ interactions. The $\mathcal O(\alpha_s)$ and the $m_b \neq 0$ corrections to the transverse-plus rate have been shown to occur only at the per-mille level in the SM~\cite{Fischer:2000kx}. Specifically, one obtains
\begin{subequations}
\begin{eqnarray}
\mathcal F_L^{\rm SM}  &=&  0.687(5)  \label{e2} \,,\\
\mathcal F_+^{\rm SM}  &=& 0.0017(1) \label{e2b}\,. 
\end{eqnarray}
\end{subequations}
One could therefore conclude that measured values of $\mathcal F_+$  exceeding $0.2\%$ level, would signal the presence of new physics (NP) beyond the SM.

When studying non-standard $tbW$ interactions, constraints from flavor changing neutral current processes involving virtual top quarks within loops play a crucial role. In particular, the inclusive decay $B\to X_s \gamma$ provides stringent bounds on the structure of $tbW$ vertices~\cite{Grzadkowski:2008mf}. One needs to take these constraints into account when evaluating the sensitivity of top decay rate measurements to potential NP contributions.

In the present paper, we study contributions of the non-SM $tbW$ interactions to the $W$ gauge boson helicity fractions in unpolarized top quark decays at next-to-leading order in QCD. We study the impact of QCD radiative corrections on NP constraints as extracted from top quark decay rate measurements and compare those with indirect bounds from inclusive radiative $B$ meson decays.

\section{Framework}
\begin{figure}[h]
\begin{center}
\includegraphics[scale=0.45]{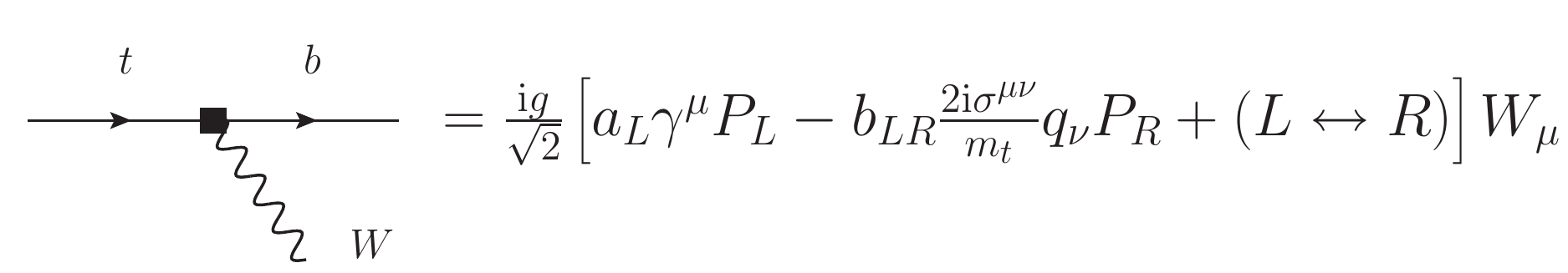}
\end{center}
\caption{\small Feynman rule for the effective $tbW$ vertex.}
\label{rule}
\end{figure}

Following \cite{AguilarSaavedra:2008zc} we work with a general effective Lagrangian for the $tbW$ interaction, which appears in the presence of new physics (NP) heavy degrees of freedom, integrated out at a scale above the top quark mass (see also~\cite{Zhang:2010px}). It can be written as
\begin{eqnarray}
{\mathcal L}_{\mathrm{eff}} = \frac{v^2}{\Lambda^2}C_L\op_{L}
+\frac{v}{\Lambda^2} C_{LR} \op_{LR} + (L \leftrightarrow R) + \mathrm{h.c.}\,,
\label{eq:Lagr}
\end{eqnarray}
with the operators defined as
\begin{eqnarray}
\op_{L} &=& \frac{g}{\sqrt{2}} W_{\mu}\Big[\bar{b}_{L}\gamma^{\mu}t_{L}\Big]\,, \label{ops} \\
\op_{LR} &=& \frac{g}{\sqrt{2}} W_{\mu\nu}\Big[\bar{b}_{L}\sigma^{\mu\nu}t_{R}\Big]\,,\nonumber
\end{eqnarray}
where $q_{R,L} = (1\pm\gamma_5)q/2$, $\sigma_{\mu\nu} = i[\gamma_\mu,\gamma_\nu]/2$ and $g$ is the weak coupling constant.
Furthermore $W_{\mu\nu} = \partial_{\mu} W_\nu - \partial_\nu W_\mu$ and $v=246$~GeV is the electroweak condensate. Finally, $\Lambda$ is the effective scale of NP. We adopt a more convenient parameterization 
\begin{eqnarray}
a_L &=& \frac{v^2}{\Lambda^2} C_L = a_L^{\mathrm{SM}}+\delta a_L= V_{tb}+\delta a_L\,,\label{def}\\
a_R &=& \frac{v^2}{\Lambda^2} C_R\,, \qquad b_{LR,RL} = \frac{v m_t}{\Lambda^2} C_{LR,RL}\nn\,,
\end{eqnarray}
resulting in the Feynman rule for the effective $tbW$ vertex as shown in Figure~\ref{rule}.
We write the complete decay width for $t\to b W$ as a sum of decay widths distinguished by different helicities of the $W$ boson
\begin{eqnarray}
\Gamma_{t\to W b}&=&\frac{m_t}{16\pi}\frac{g^2}{2}\sum_{i}\Gamma^i\,,
\end{eqnarray}
where $i= L,+,-$ stands for longitudinal, transverse-plus and transverse-minus.

The $\Gamma^i$ decay rates have already been studied to quite some extent in the existing literature. The tree-level analysis of the effective interactions in (\ref{eq:Lagr}) has been conducted in Ref.~\cite{AguilarSaavedra:2006fy}. QCD corrections, however, have been studied only for the chirality conserving operators. Results for a general parametrization can be found in Ref.~\cite{Bernreuther:2003xj}, while SM analysis is given in \cite{Fischer:2001gp, Czarnecki:2010gb}, where ${\cal O}(\alpha_s)$ results including $m_b\neq 0$ effects and ${\cal O}(\alpha_s^2)$, $m_b=0$ corrections have been computed. The hard gluon emission corrections are especially important for the observable $\mathcal F_+$ since they allow to lift the helicity  suppression present at the leading order (LO) in the SM. Helicity suppression in this observable is also exhibited in the presence of  the NP operator  $\mathcal O_{LR}$, which is especially interesting since it is least constrained by indirect bounds coming from the $B\to X_s \gamma$ decay rate~\cite{Grzadkowski:2008mf} and thus has the potential to modify the $t\to bW$ decay properties in an observable way.

\section{Results}
\begin{figure}[h!]
\begin{center}
\includegraphics[scale=0.5]{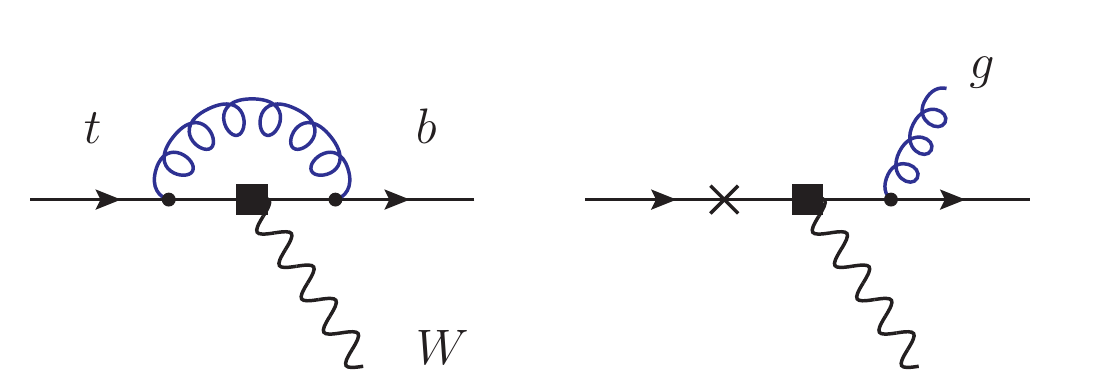}
\end{center}
\caption{Diagrams for next-to-leading order QCD contributions. Cross marks the additional points from which the gluon can be emitted. }
\label{feyndiags}
\end{figure}

We compute the ${\cal O}(\alpha_s)$ corrections to the polarized rates $\Gamma^i$ in the $m_b=0$ limit including both operators given in eq.~(\ref{ops}) (and their chirality flipped counterparts). The appropriate Feynman diagrams are presented in Figure~\ref{feyndiags}. We regulate UV and IR divergences by working in $d=4+\epsilon$ dimensions. The renormalization procedure closely resembles the one described in \cite{Drobnak:2010by}. To avoid conceivable problems regarding $\gamma^5$ in $d$-dimensions, we use the prescription of Ref.~\cite{Larin:1993tq}. To project out the desired helicities of the $W$ boson we use the technique of covariant projectors as described by Fischer et al. in Ref.~\cite{Fischer:2000kx}.

\subsection{The decay rates}
In the $m_b=0$ limit there is no mixing between chirality flipped operators and the decay rates can be written as
\begin{eqnarray}
\Gamma^{(L,+,-)}&=& |a_L|^2 \Gamma^{(L,+,-)}_a + |b_{LR}|^2 \Gamma^{(L,+,-)}_{b} \label{e3}\\
&+& 2\mathrm{Re}\{a_L b_{LR}^*\} \Gamma^{(L,+,-)}_{ab} + {\scriptstyle\langle L \leftrightarrow R,+\leftrightarrow -\rangle}\,.\nonumber
\end{eqnarray}
Analytical formulae for $\Gamma^{i}_{a,b,ab}$ functions are given in the appendix. We have crosschecked $\Gamma^i_{a}$ with the corresponding expressions given in \cite{Fischer:2000kx} and found agreement between the results.
\begin{table}[h!]
\begin{tabular}{l|c|c|c||c}
			& $L$ 	& $+$ 	& $-$&unpolarized 	 \\\hline
$\Gamma_a^{i,\mathrm{LO}}$&$\frac{(1-x^2)^2}{2x^2}$& $0$ & $(1-x^2)^2$&$\frac{(1-x^2)^2(1+2x^2)}{2x^2}$\\
$\Gamma_b^{i,\mathrm{LO}}$& $2x^2(1-x^2)^2$ &$0$&$4(1-x^2)^2$&$2(1-x^2)^2(2+x^2)$\\
$\Gamma_{ab}^{i,\mathrm{LO}}$&$(1-x^2)^2$&$0$&$2(1-x^2)^2$&$3(1-x^2)^2$\\
\end{tabular}
\caption{\small Tree-level decay widths for different $W$ helicities and the their sum, which gives the unpolarized width. All results are in the $m_b=0$ limit and we have defined $x=m_W/m_t$.}
\label{tab1}
\end{table}
The LO (${\cal O}(\alpha_s^0)$) contributions to decay rates $\Gamma_{a,b,ab}^{i,\mathrm{LO}}$ are obtained with a tree-level calculation and are given in Table~\ref{tab1}. Our results coincide with those given in \cite{AguilarSaavedra:2006fy}, if the mass $m_b$ is set to zero. 
\begin{table}[h]
\begin{tabular}{c|c|c|c}
 &$L$ & $+$ &$-$\\\hline
$\Gamma^{i,\mathrm{NLO}}_a/\Gamma_a^{i,\mathrm{LO}}$  &$0.90$&$3.50$&$0.93$\\
$\Gamma^{i,\mathrm{NLO}}_b/\Gamma_b^{i,\mathrm{LO}}$  &$0.96$&$4.71$&$0.91$\\
$\Gamma^{i,\mathrm{NLO}}_{ab}/\Gamma_{ab}^{i,\mathrm{LO}}$ &$0.93$&$3.75$&$0.92$
\end{tabular}
\caption{\small Numerical values for $\Gamma^\mathrm{NLO}/\Gamma^{\mathrm{LO}}$ with the following input parameters $m_t=173$ GeV, $m_W = 80.4$ GeV, $\alpha_s(m_t) = 0.108$. Scale $\mu$ appearing in NLO expressions is set to $\mu=m_t$. In addition $m_b=4.8$ GeV. These values are used throughout the paper for all numerical analysis. }\label{LO/NLO}
\end{table}
The change of $\Gamma_{a,b,ab}^{i}$ going form LO to next-to-leading order (NLO) in $\alpha_s$ is presented in Table~\ref{LO/NLO}. Since in the $m_b=0$ limit $\Gamma_{a,b,ab}^{+,\mathrm{LO}}$ vanish, we use the full $m_b$ dependence of the LO rate when dealing with $W$ transverse-plus helicity. Effectively we neglect the ${\cal O}(\alpha_s m_b)$ contributions. 
In Ref.~\cite{Fischer:2001gp} it has been shown, that these subleading contributions can scale as $\alpha_s (m_b/m_W)^2 \log (m_b/m_t)^2 $ leading to a relative effect of a couple of percent compared to the size of $\mathcal O (\alpha_s)$ corrections in the $m_b=0$ limit.

\subsection{Effects on ${\cal F}_+$}
\begin{figure}[h]
\begin{center}
\includegraphics[scale=0.6]{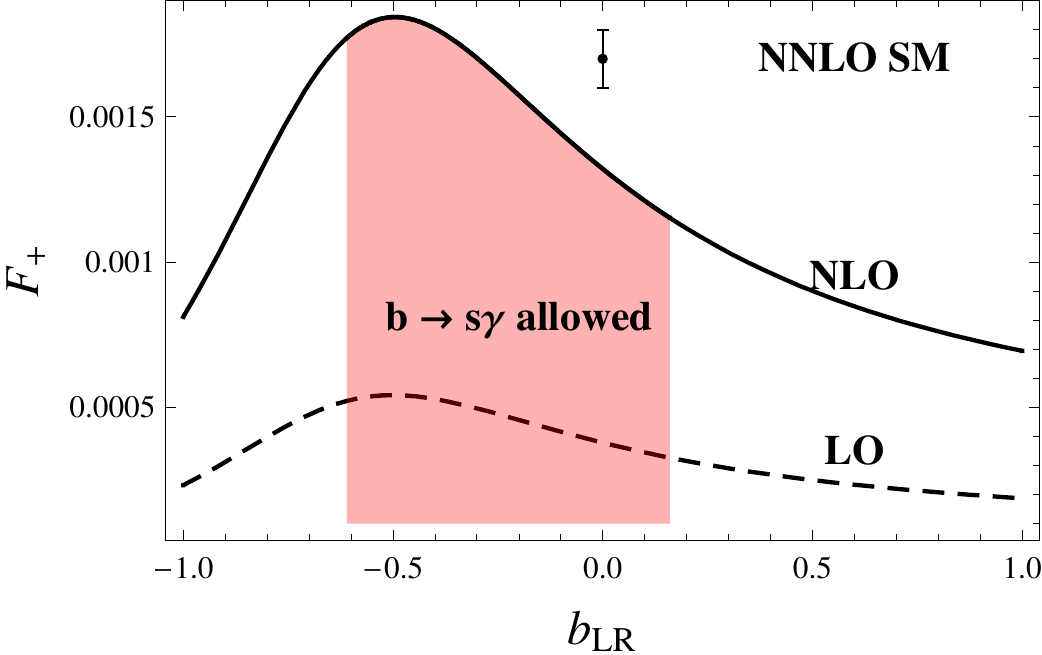}
\end{center}
\caption{\small Value of ${\cal F}_+$ as a function of $b_{LR}$ (other NP coefficients being set to zero). Red band shows the allowed interval for $b_{LR}$ as given in Ref.~\cite{Grzadkowski:2008mf}. Dashed line coresponds to LO results at $m_b\neq 0$, while the solid line represents the ${\cal O}(\alpha_s)$ results. We also present the SM ${\cal O}(\alpha_s^2)$ (NNLO) value given in eq.~(\ref{e2b}).}
\label{bLR1}
\end{figure}

We have analyzed the effects on ${\cal F}_+$ when going from LO to NLO in QCD. Assuming the NP coupling parameters to be real, we consider contributions of a single NP operator at a time. Present $95\%$ C.L. constrains on $\delta a_L,a_R,b_{LR},b_{RL}$ come from the weak radiative $B$ meson decays ($b \to s \gamma$) analyzed in Ref.~\cite{Grzadkowski:2008mf}. Translated to our definition of parameters these bounds read
\begin{align}
 -0.13 \le \delta a_L \le 0.03\,, & -0.0007 \le  a_R \le 0.0025\,,\label{mejeOLD} \\
 -0.61 \le b_{LR} \le 0.16\,, & -0.0004 \le b_{RL} \le 0.0016\,.\nonumber
\end{align}
Compared with others, constraints on $b_{LR}$ are considerably looser. We present the effect of $b_{LR}$ on ${\cal F}_+$ in Figure~\ref{bLR1}. We see that the the increase is substantial when going to NLO in QCD, but still leaves ${\cal F}_+$ at the $1-2$ per-mille level.
We summarize the effects of the other NP operators on $\mathcal F_+$ in Table~\ref{tab3}. Nonstandard value of $a_L$ does not effect the different $W$ helicity branching fractions which are the same as in the SM. The  dependence of ${\cal F}_+$ on non-zero values of $a_R$ and $b_{RL}$ in the $b\to s \gamma$ allowed region is mild, reaching a maximum at the lowest allowed values of $a_R$ and $b_{RL}$.
\begin{table}[h]
\begin{tabular}{c|c|c|c}
& SM ($\delta a_L$) & $a_R$ & $b_{RL}$\\\hline
${\cal F}_+^{\mathrm{NLO}}/{\cal F}_+^{\mathrm{LO}}$&$3.49$ & $3.40$& $3.38$\\
${\cal F}_+^{\mathrm{NLO}}/10^{-3}$&$1.32 $ &$1.34$ &$1.34$
\end{tabular}
\caption{\small Maximum allowed effects on ${\cal F}_+$ due to non-zero values of $a_R$ and $b_{RL}$ at ${\cal O(\alpha_s)}$.}\label{tab3}
\end{table}
We observe that for these NP contributions, $b\to s \gamma$ already constrains the value of ${\cal F}_+$ to be within $2\%$ of the SM prediction.

\subsection{Effects on ${\cal F}_L$}
\begin{figure}[!h]
\begin{center}
\includegraphics[scale=0.6]{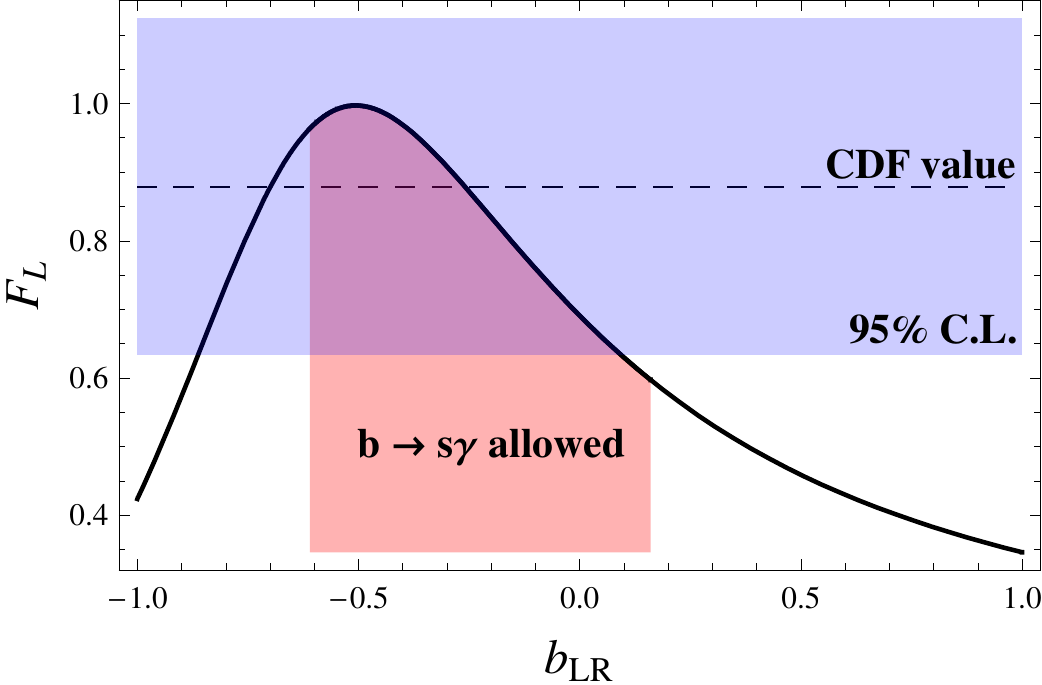}
\end{center}
\caption{\small Value of ${\cal F}_L$ as a function of $b_{LR}$ (other NP coefficients being set to zero). Red band shows the allowed interval for $b_{LR}$ as given in Ref.~\cite{Grzadkowski:2008mf}. We also present the CDF values given in eq.~(\ref{e1}). }
\label{bLRlimit}
\end{figure}

Analyzing a single real NP operator contribution at the time, leading QCD corrections decrease ${\cal F}_L$ by approximately $1\%$ in all cases. Possible effects of $a_R$ and $b_{RL}$ are again severely constrained by $b\to s \gamma$. On the other hand, we find that the most recent CDF measurement of $\mathcal F_L$ in eq.~(\ref{e1}) already allows to put competitive bounds on $b_{LR}$ compared to the indirect constraints given in eq.~(\ref{mejeOLD}). We plot the dependence of ${\cal F}_L$ on $b_{LR}$ in Figure~\ref{bLRlimit}. A new $95\%$ C.L. upper bound is found to be
\begin{eqnarray}
b_{LR} < 0.09 \,, \hspace{0.2cm} 95\%\,\,\text{C.L.}\,.\label{newlimit}
\end{eqnarray}

\section{Conclusions}
We have analyzed the decay of an unpolarized top quark to a bottom quark and a polarized $W$ boson as mediated by the most general effective $tbW$ vertex at ${\cal O}(\alpha_s)$. We have shown, that within this approach the helicity fraction ${\cal F}_+$  can reach maximum values of the order of $2$ per-mille in the presence of a non-SM effective operator $\mathcal O_{LR}$. Leading QCD effects increase the contributions of $\mathcal O_{LR}$ substantially owing to the helicity suppression of the LO result, while other considered NP effective operator contributions are much less affected. Indirect constraints coming from the $B\to X_s \gamma$ decay rate already severely restrict the contributions of these NP operators. In particular, considering only real contributions of a single NP operator at a time, all considered operators except $\mathcal O_{LR}$ are constrained to yield $\mathcal F_+$ within $2\%$ of the SM prediction.  Even in the presence of the much less constrained $\mathcal O_{LR}$ contributions,  a potential determination of ${\cal F}_+ $ significantly deviating from the SM prediction, at the projected sensitivity of the LHC experiments~\cite{AguilarSaavedra:2007rs}, could not be explained within such framework. Based on the existing SM calculations of higher order QCD and electroweak corrections \cite{Czarnecki:2010gb, Do:2002ky}, we do not expect such corrections to significantly affect our conclusions.

Finally, we have set a new $95\%$ C.L. upper bound on the $b_{LR}$ contributions given in eq.~(\ref{newlimit}), lowering the previous indirect bound coming from $B\to X_s \gamma$ decay by $44\%$. With increased precision of the ${\cal F}_L $ measurements at the Tevatron and the LHC this bound (as well as the lower bound on the same coupling) is expected to be further significantly improved in the near future.

\begin{acknowledgments}
This work is supported in part by the European Commission RTN  network, Contract No. MRTN-CT-2006-035482 (FLAVIAnet) and by the Slovenian Research Agency. 
\end{acknowledgments}

\appendix
\begin{widetext}
\newpage
\section{Analytical Formulae}
In this appendix we present analytical formulae for all nine $\Gamma^{L,+,-}_{a,b,ab}$ appearing in eq.~(\ref{e3}) to ${\cal 
O}(\alpha_s)$ order and in the $m_b=0$ limit. Here $\mu$ is the arbitrary scale, remnant of operator renormalization, $x= m_W/m_t$ and $C_F=4/3$. For the computation of ${\cal F}_i$, we have also used the $\Gamma_{a,b,ab}$, summed over the three $W$ helicity states. We omit these expressions here, as they coincide with the analoge formulae given in  \cite{Drobnak:2010wh,Drobnak:2010by} obtained in the context of $t\to c Z$ decays.
\vspace{-0.5cm}
\subsection{Longitudinal polarization}
\vspace{-0.8cm}
\begin{eqnarray}
\Gamma_{a}^L &=&\frac{(1-x^2)^2}{2 x^2}+\frac{\alpha_s}{4\pi}C_F\Bigg[ \frac{(1-x^2)(5+47 x^2-4x^4)}{2 x^2}-\frac{2\pi^2}{3}\frac{1+5x^2+2x^4}{x^2}-\frac{3(1-x^2)^2}{x^2}\log(1-x^2)\nn\\
&-&\frac{2(1-x)^2(2-x+6x^2+x^3)}{x^2}\log(x)\log(1-x)- \frac{2(1+x)^2(2+x+6x^2-x^3)}{x^2}\log(x)\log(1+x)\nn\\
&-&\frac{2(1-x)^2(4+3x+8x^2+x^3)}{x^2}\mathrm{Li}_2(x)-\frac{2(1+x)^2(4-3x+8x^2-x^3)}{x^2}\mathrm{Li}_2(-x)+16(1+2x^2)\log(x)\Bigg]\,,\\
\Gamma_b^L&=&2x^2(1-x^2)^2+\frac{\alpha_s}{4\pi}C_F\Bigg[-2x^2(1-x^2)(21-x^2)+\frac{2\pi^2}{3}4x^2(1+x^2)(3-x^2)+4x^2(1-x^2)^2\log\Big(\frac{m_t^2}{\mu^2}\Big)\nn\\
&-&16x^2(3+3x^2-x^4)\log(x)-4(1-x^2)^2(2+x^2)\log(1-x^2)-8x(1-x)^2(3+3x^2+2x^3)\log(x)\log(1-x)\nn\\
&+&8x(1+x)^2(3+3x^2-2x^3)\log(x)\log(1+x)-8x(1-x)^2(3+2x+7x^2+4x^3)\mathrm{Li}_2(x)\nn\\
&+&8x(1+x)^2(3-2x+7x^2-4x^3)\mathrm{Li}_2(-x)\Bigg]\,,\\
\Gamma_{ab}^L&=&(1-x^2)^2+\frac{\alpha_s}{4\pi}C_F\Bigg[-(1-x^2)(1+11x^2)-\frac{2\pi^2}{3}(1-7x^2+2x^4)+(1-x^2)^2 \log\Big(\frac{m_t^2}{\mu^2}\Big)\nn\\
&-&\frac{2(1-x^2)^2(1+2x^2)}{x^2}\log(1-x^2)-4x^2(7-x^2)\log(x)-4(1-x)^2(1+5x+2x^2)\log(x)\log(1-x)\nn\\
&-&4(1+x)^2(1-5x+2x^2)\log(x)\log(1+x)-4(1-x)^2(3+9x+4x^2)\mathrm{Li}_2(x)-4(1+x)^2(3-9x+4x^2)\mathrm{Li}_2(-x)\Bigg]\,.\nn\\
\end{eqnarray}
\vspace{-1cm}
\subsection{Transverse-plus polarization}
\vspace{-1cm}
\begin{eqnarray}
\Gamma_{a}^+ &=&\frac{\alpha_s}{4\pi}C_F\Bigg[-\frac{1}{2}(1-x)(25+5x+9x^2+x^3)+\frac{\pi^2}{3}(7+6x^2-2x^4)-2(5-7x^2+2x^4)\log(1+x)\nn\\
&-&2(5+7x^2-2x^4)\log(x)-\frac{(1-x)^2(5+7x^2+4x^3)}{x}\log(x)\log(1-x)-\frac{(1-x)^2(5+7x^2+4x^3)}{x}\mathrm{Li}_2(x)\nn\\
&+&\frac{(1+x)^2(5+7x^2-4x^3)}{x}\log(x)\log(1+x)+\frac{5+10x+12x^2+30x^3-x^4-12x^5}{x}\mathrm{Li}_2(-x)\Bigg]\,,\\
\Gamma_{b}^+ &=&\frac{\alpha_s}{4\pi}C_F\Bigg[ \frac{4}{3}x(1-x)(30+3x+7x^2-2x^3-2x^4)-4\pi^2 x^4-8(5-9x^2+4x^4)\log(1+x)+8x^2(1+5x^2)\log(x)\nn\\
&-&4(1-x)^2(4+5x+6x^2+x^3)\log(x)\log(1-x)-4(1+x)^2(4-5x+6x^2-x^3)\log(x)\log(1+x)\nn\\
&-&4(1-x)^2(4+5x+6x^2+x^3)\mathrm{Li}_2(x)-4(4+3x-16x^2+6x^3+16x^4-x^5)\mathrm{Li}_2(-x)\Bigg]\,,\\
\Gamma_{ab}^+ &=&\frac{\alpha_s}{4\pi}C_F\Bigg[
2x(1-x)(15-11x)+\frac{2\pi^2}{3}x^2(5-2x^2)-2(13-16x^2+3x^4)\log(1+x)+2x^2(1+3x^2)\log(x)\nn\\
&-&2(1-x)^2(5+7x+4x^2)\log(x)\log(1-x)-2(1+x)^2(5-7x+4x^2)\log(x)\log(1+x)\nn\\
&-&2(1-x)^2(5+7x+4x^2)\mathrm{Li}_2(x)-2(3+3x-31x^2+x^3+12x^4)\mathrm{Li}_2(-x)\Bigg]\,.
\end{eqnarray}

\subsection{Transverse-minus polarization}
\vspace{-1cm}
\begin{eqnarray}
\Gamma_{a}^- &=&(1-x^2)^2+\frac{\alpha_s}{4\pi}C_F\Bigg[-\frac{1}{2}(1-x)(13+33x-7x^2+x^3)+\frac{\pi^2}{3}(3+4x^2-2x^4)-2(5+7x^2-2x^4)\log(x)\nn\\
\nn&-&\frac{2(1-x^2)^2(1+2x^2)}{x^2}\log(1-x)-\frac{2(1-x^2)(1-4x^2)}{x^2}\log(1+x)-\frac{(1-x)^2(5+7x^2+4x^3)}{x}\log(x)\log(1-x)\\
\nn&+&\frac{(1+x)^2(5+7x^2-4x^3)}{x}\log(x)\log(1+x)-\frac{(1-x)^2(5+3x)(1+x+4x^2)}{x}\mathrm{Li}_2(x)\\
&+&\frac{5+2x+12x^2+6x^3-x^4-4x^5}{x}\mathrm{Li}_2(-x)\Bigg]\,,\\
\Gamma_{b}^- &=&4(1-x^2)^2+\frac{\alpha_s}{4\pi}C_F\Bigg[\frac{4}{3}(1-x)(16-14x+22x^2+18x^3-3x^4-3x^5)\nn\\
\nn&-&\frac{\pi^2}{3}4(4+x^4) +8x^2(1+5x^2)\log(x)-24(1-x^2)^2\log(1-x)+8(1-x^2)(2-x^2)\log(1+x)\\
\nn&-&4(1-x)^2(4+5x+6x^2+x^3)\log(x)\log(1-x)-4(1+x)^2(4-5x+6x^2-x^3)\log(x)\log(1+x)\\
&-&4(1-x)^2(12+21x+14x^2+x^3)\mathrm{Li}_2(x)-4(12+3x+6 x^3-x^5)\mathrm{Li}_2(-x)+8(1-x^2)^2\log\Big(\frac{m_t^2}{\mu^2}\Big)\Bigg]\,,\\
\Gamma_{ab}^- &=&2(1-x^2)^2+\frac{\alpha_s}{4\pi}C_F\Bigg[2(1-x)(9-6x+6x^2-5x^3)-\frac{2\pi^2}{3}(5+2x^4)+2x^2(1+3x^2)\log(x)\nn\\
\nn&-&\frac{2(1-x^2)^2(1+5x^2)}{x^2}\log(1-x)-\frac{2(1-x^2)(1-9x^2-2x^4)}{x^2}\log(1+x)\\
\nn&-&2(1-x)^2(5+7x+4x^2)\log(x)\log(1-x)-2(1+x)^2(5-7x+4x^2)\log(x)\log(1+x)\\
&-&2(1-x)^2(13+23x+12x^2)\mathrm{Li}_2(x)-2(15+3x+5x^2+x^3+4x^4)\mathrm{Li}_2(-x)+2(1-x^2)^2\log\Big(\frac{m_t^2}{\mu^2}\Big)\Bigg]\,.
\end{eqnarray}
\end{widetext}

\end{document}